\documentclass[]{aa}  

\usepackage{graphicx}
\usepackage{hyperref}
\usepackage{txfonts}
\usepackage{subfigure}
\usepackage[switch]{lineno}
\def\msol{\hbox{\kern 0.20em $M_\odot$}}
\def\lsol{\hbox{\kern 0.20em $L_\odot$}}
\def\rsol{\hbox{\kern 0.20em $R_\odot$}}
\def\sr{\hbox{\kern 0.20em sr}}
\def\srmu{\hbox{\kern 0.20em sr$^{-1}$}}
\def\g{\hbox{\kern 0.20em g}}
\def\gmu{\hbox{\kern 0.20em g$^{-1}$}}
\def\kg{\hbox{\kern 0.20em kg}}
\def\pc{\hbox{\kern 0.20em pc}}
\def\mum{\hbox{\kern 0.20em $\mu$m}}
\def\mumd{\hbox{\kern 0.20em $\mu$m$^{-2}$}}
\def\cm{\hbox{\kern 0.20em cm}}
\def\m{\hbox{\kern 0.20em m}}
\def\km{\hbox{\kern 0.20em km}}
\def\nm{\hbox{\kern 0.20em nm}}
\def\s{\hbox{\kern 0.20em s}}
\def\h{\hbox{\kern 0.20em h}}
\def\sec{\hbox{\kern 0.20em sec}}
\def\min{\hbox {\kern 0.20em min}}
\def\smu{\hbox{\kern 0.20em s$^{-1}$}}
\def\smd{\hbox{\kern 0.20em s$^{-2}$}}
\def\an{\hbox{\kern 0.20em an}}
\def\anmu{\hbox{\kern 0.20em an$^{-1}$}}
\def\deg{\hbox{\kern 0.20em $^{\rm o}$}}
\def\yr{\hbox{\kern 0.20em yr}}
\def\yrmu{\hbox{\kern 0.20em yr$^{-1}$}}
\def\Myr{\hbox{\kern 0.20em Myr}}
\def\Mymu{\hbox{\kern 0.20em Myr$^{-1}$}}
\def\K{\hbox{\kern 0.20em K}}
\def\mK{\hbox{\kern 0.20em mK}}
\def\pcmu{\hbox{\kern 0.20em pc$^{-1}$}}
\def\pcmd{\hbox{\kern 0.20em pc$^{-2}$}}
\def\pcmt{\hbox{\kern 0.20em pc$^{-3}$}}
\def\kms{\hbox{\kern 0.20em km\kern 0.20em s$^{-1}$}}
\def\kmpd{\hbox{\kern 0.20em km$^{2}$}}
\def\kpc{\hbox{\kern 0.20em kpc}}
\def\cms{\hbox{\kern 0.20em cm\kern 0.20em s$^{-1}$}}
\def\erg{\hbox{\kern 0.20em erg}}
\def\ergs{\hbox{\kern 0.20em erg}}
\def\cmpd{\hbox{\kern 0.20em cm$^2$}}
\def\cmmd{\hbox{\kern 0.20em cm$^{-2}$}}
\def\cmms{\hbox{\kern 0.20em cm$^{-6}$}}
\def\cmpt{\hbox{\kern 0.20em cm$^3$}}
\def\cmmt{\hbox{\kern 0.20em cm$^{-3}$}}
\def\mpd{\hbox{\kern 0.20em m$^2$}}
\def\mmd{\hbox{\kern 0.20em m$^{-2}$}}
\def\mpt{\hbox{\kern 0.20em m$^3$}}
\def\mmt{\hbox{\kern 0.20em m$^{-3}$}}
\def\mujy{\hbox{\kern 0.20em $\mu$Jy}}
\def\mjy{\hbox{\kern 0.20em mJy}}
\def\Mj{\hbox{\kern 0.20em MJy}}
\def\jy{\hbox{\kern 0.20em Jy}}
\def\ghz{\hbox{\kern 0.20em GHz}}
\def\srmd{\hbox{\kern 0.20em sr$^{-1}$}}

\def \mum{$\mu$m}
\def\G{\hbox{\kern 0.20em G}}

\def\h13cop{\hbox{H$^{13}$CO$^{+}$}}

\def\h2o{\hbox{H$_2$O}}

\begin{document} 
\nolinenumbers

\title{Outflow Interaction in Cep-E: }

   \subtitle{Numerical Simulation and Observational Manifestation}

   \author{        
          A. Rodríguez-González \inst{1,3}
          \and
       P. R. Rivera-Ortiz\inst{2,1}
       \and
       Z. Meliani \inst{3}
       \and
       E. Alquicira-Peláez \inst{1} 
       \and
       A. Durán \inst{2} 
       \and
       Luis A. Zapata \inst{2} 
          }

   \institute{
Instituto de Ciencias Nucleares, Universidad Nacional Aut\'onoma de M\'exico, Ap. 70-543, 04510, Ciudad de M\'exico, M\'exico.
\and 
   Instituto de Radioastronom\'ia y Astrof\'isica, Universidad Nacional Aut\'onoma de M\'exico, 58090, Morelia, Michoac\'an, Mexico
    \and
    LUTH, Observatoire de Paris, CNRS, PSL, Universit\'{e} de Paris; 5 Place Jules Janssen, 92190 Meudon, France
    }

\date{Received  April 1st, 2023; accepted XXXX, 2023}


\abstract
   {There is clear observational evidence that the main Class 0/I stages of the star formation process are associated with powerful collimated outflows (jets), which sometimes propagate up to distances as large as  $10^{4-5}$ au scales in molecular clouds. Additionally, intermediate high-mass and low-mass protostars have often been observed to form in crowded clusters, where the typical separation distance between any two cluster members is of the same order or smaller than the scale of the outflow length.   Therefore, there must be an interaction between the molecular outflows of different protostars within the protostellar association. A good example of this is the case of Cepheus E-mm, which is a protostellar outflow extending over a few dozen au. At its core is a binary system consisting of two protostars, Cep E-A and Cep E-B, separated by about 1000 au. Both protostars eject molecular jets at velocities of ~100 km/s. The interaction between these molecular outflows provides an opportunity to study the effects of jet collisions in a clustered star-forming environment, as they may leave detectable imprints on the morphology of the main envelope of the system.
   }
   { Our work aims to study the effects of the collision of molecular jets associated with the components of the binary system Cep-A and Cep-E, analyzing the disruption or reduction of molecular emission in the main envelope of the system, which the molecular outflow { launched} by Cep-A presumably pushes. If we characterize the collision in this system, we can provide insights into the expected morphology and molecular emissions in collisions of molecular outflows { associated to star forming process.}
{ To achieve this, we have developed numerical models based on the observed physical conditions of Cep E–South, which may lead to a strong interaction between molecular outflows capable of altering the evolution of this object}. We will also search for the expected numerical signatures at length scales accessible to the (sub)millimeter interferometers ALMA and NOEMA in the largest outflows.}
   {We have used the 3D hydrodynamical code to model a system of two protostellar jets powered by two sources interacting in a configuration similar to the prototypical intermediate-mass protobinary system Cep\,E. Maps of density, emissivity, and velocity distribution have been produced, allowing direct comparison with these observations. }
   { We have found signatures of outflow interaction that can be used to find more interaction cases.}
   {}


   \maketitle
%

\section{Introduction}
 {Hypersonic collimated jets, surrounded by a lower-velocity outflow, are commonly observed in Class 0/I objects, corresponding to the earliest evolutionary phases of protostars  \citep{Ray2021}}. In {the past} decades, understanding the formation of protostellar outflows, the role of jets, and their interaction with the ambient medium has been a real challenge. { Progress in  2D or 3D hydrodynamical simulations \citep{Rabenanahary2022,Rivera-Ortiz2023} have brought convincing support, both morphologically and dynamically, to the model of jet-driven bow shock mechanism initially proposed by \cite{Raga-Cabrit1993}. These models consider isolated forming protostars, whose jets may be morphologically modified by the interaction of a binary component that adds a velocity component tangent to the orbital motion \citep{Schutzer2022} or precession \citep{Rivera-Ortiz2023} as a possible consequence of the interaction of binary systems.}

However, source multiplicity in protostellar systems appears to be very common \cite[see, e.g. for a review][]{Reipurth2014}. For instance,  
the SMA study by \citet{Chen2013}, a sample of 33 Class 0 protostars, all located less than 400 pc away,  showed that more than 2/3 of the source sample displayed evidence of a companion at a distance of 50--5000 au. Similar results were obtained consistently towards e.g. BHR71 or Cep\,E, IRAS4A \citep{Zapata2018,Ospina-Zamudio2018,Taquet2020,Toledano2023}.
Hence, the possibility that several outflows could propagate inside a single parental protostellar envelope and interact with the ambient gas or other outflows can no longer be discarded or considered a low-probability phenomenon. 

As intermediate to high-mass stars form in dense clusters (e.g., Orion with $\sim$106 stars pc$^{-3}$, \cite{RivillaETAL2014}), dynamical interactions among cluster members may be very common \citep{BonellBate2005}. In the case of {massive} protostars with circumstellar disks and envelopes sizes of a {few thousand astronomical units}, their impact areas should increase dramatically, producing frequent interactions at the centre of the cluster. The rich cluster of outflows detected in the CO 2-1 and SiO 5-4 lines in the W43-MM1 high-mass star forming region by \cite{Nony2020,Nony2023} {(see their Figure~1)}, as part of the ALMA-IMF Large Program, suggests that multiplicity is present in the 27 protostellar cores of the cluster, with multiple mass-ejection phenomena interacting at $5\times10^3$ au scale.  Recently, observations at $10^3$~au scale have become commonly accessible with the large (sub)millimeter facilities ALMA and  NOEMA only recently, the new instruments are revealing faint structures within the outflow, so it is only now that observational constraints can be brought on flow/jet interactions from multiple sources in a common protostellar core.  

{ The present study aims to shed light on the physical processes involved in the collision of outflow/jet envelopes in protostellar binary systems and to propose observational signatures that could provide empirical evidence for such events. However, numerous free parameters—such as the physical properties of the outflows, variability, and the angle between interacting outflows—depend solely on the specific object under study. Therefore, we focus on the interaction between the molecular outflows ejected by the Cep-A and Cep-B components in the Cep E-mm system. Cep-A and Cep-B form a binary protostellar system, separated by approximately 1000 au, with supersonic molecular outflows with velocities in the order of $\sim$100 km/s \citep{LeFloch1996,Ospina-Zamudio2018,Schutzer2022}. The collision of these molecular outflows could be reflected in the morphology of the system structure.

Using a uniform-mesh three-dimensional hydrodynamical code \citep{Rodriguez-Gonzalez2019}, which accounts for the cooling rate at low temperatures, we have modelled the collision of the jets powered by these forming protostars within a dense and stratified protostellar envelope. We are adopting initial conditions for the jet and the parental core based on the observational studies of Class 0 protostellar core Cep\,E-mm  \citep{Schutzer2022,Lefloch2015,Crimier2010}, which can be considered as a prototype for such observational and numerical studies \citep{Rivera-Ortiz2023}.}

Our article is organized as follows. Section \ref{sec:Sec2} briefly describes the astrophysical source used to determine the initial jet and protostellar conditions in the numerical simulations. Section \ref{Sec:3} describes the hydrodynamical code  used to carry out the numerical simulations 
and the physical constraints we adopted for the modelling based on the CepE observations. The results of our simulations are discussed in Section \ref{Sec:4};  we first show how the dynamics of both jets and their outflows can be strongly perturbed under specific jet dynamical conditions, both at small and large scales. Interestingly, the numerical signatures of the interaction are consistent with the observational study of \citet{Schutzer2022}.  We also present the various observational signatures to be expected based on preliminary exploration of the outflow corresponding parameter space.  Finally, our conclusions are presented in Section \ref{Sec:5}.

%

\section{Cepheus E}
\label{sec:Sec2}
\begin{figure}
\centering
\includegraphics[trim={2.5cm 14cm 2.5cm 3cm},width=1.1\columnwidth]{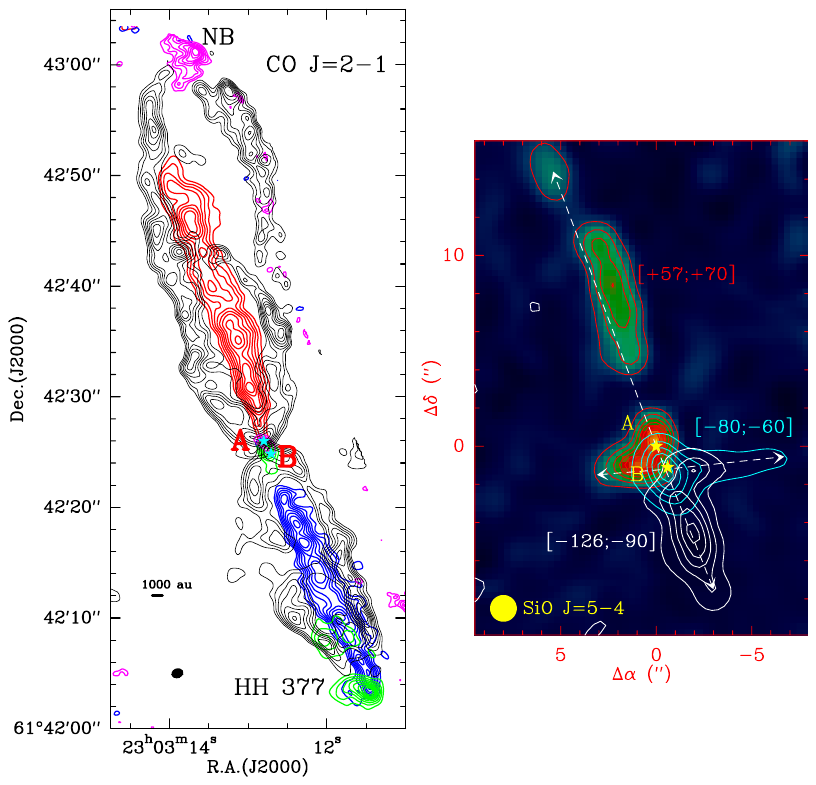}
\caption{\label{fig01}(Left) CO $J$= 2--1 emission detected in the Cep\,E protostellar core with the IRAM interferometer at 1\arcsec (820 au) resolution. The black contours trace the emission from the low-velocity gas ($|V-V_{lsr}|$ < $8\kms$), and the blue/red contours trace the high-velocity jet ($V-V_{lsr}$ > $50\kms$, powered by component A of the Cep\,E protostellar core. The data are from \citet{Lefloch2015} and \citet{Schutzer2022}. (Right)~SiO $J$= 5--4 emissions detected in the Cep\,E protostellar core with the IRAM interferometer at $1.4\arcsec$ (1150 au) resolution in the jets powered by protostars A (red and white contours) and B (blue and red contours). The velocity intervals of integrated flux are reported (in km~s$^{-1}$) next to the lobes of the jets from A and B.  The first contour and contour interval are $10\%$ and $20\%$ of the maximum peak intensity. The data are from \citet{Ospina-Zamudio2018}.}
\end{figure}


Cep\,E-mm is the prototype of an intermediate-mass star forming core, located in the Cepheus molecular cloud at a distance of $d=820\pc$ \citep{Schutzer2022}, with a bolometric luminosity of $L_{\rm b}\sim 100$L$_\sun$ \citep{LeFloch1996}. Its millimeter emission peak is centered at  $\alpha$(J2000)= $23^h03^m12.8^s$, $\delta$(J2000)= $+61\deg 42\arcmin 26\arcsec$. The  Cep\,E-mm core is a binary system of  two protostars Cep\,E-A and Cep\,E-B,  according to \citet{Ospina-Zamudio2018} they have a separation of $1.35\arcsec$ (1000 au), which both power molecular jets with velocities up to 80--120$\kms$ detected in the millimetre lines of various molecular species such as CO,  SiO, as can be seen in 
Figure~\ref{fig01}  from \citet{Ospina-Zamudio2018}. { They also determined the physical properties of components A and B. Based on 1.3 mm continuum observations, they showed that CepE-A has a size of $875\,\text{au}$, a mass of $0.56\,M_{\odot}$, a 1.3 mm total flux of $169$ mJy, and a dust temperature of $T_d = 60\,\text{K}$, while CepE-B has a size of $670\,\text{AU}$, a mass of $0.13\,M_{\odot}$, a 1.3 mm total flux of $40$ mJy, and the same dust temperature of $T_d = 60\,\text{K}$. }

The main axis of jets Cep\,E-A and Cep\,E-B are close to $90\deg$, almost perpendicular, one to the other.
As can be seen in Figure~\ref{fig01}, the high-velocity jet powered by source A propagates at (radial) velocities of about $+65\kms$ and $-110\kms$ in the Northern (red contours in Figure~\ref{fig01}) and Southern (blue contours in Figure \ref{fig01}) lobes, respectively, regarding the ambient cloud gas ($V_{lsr}$=-$10.9\kms$). In comparison, source B and its jet have been much less studied. { Still, the jet radial velocities are on the same order of magnitude as those observed in the jet from source A} with typical values of $+/-70\kms$ in both red- and blueshifted lobes regarding the ambient cloud (right panel in Figure~\ref{fig01}).

\begin{figure}
\includegraphics[width=0.9\columnwidth,trim=0 0 7.75cm 0,clip=true]{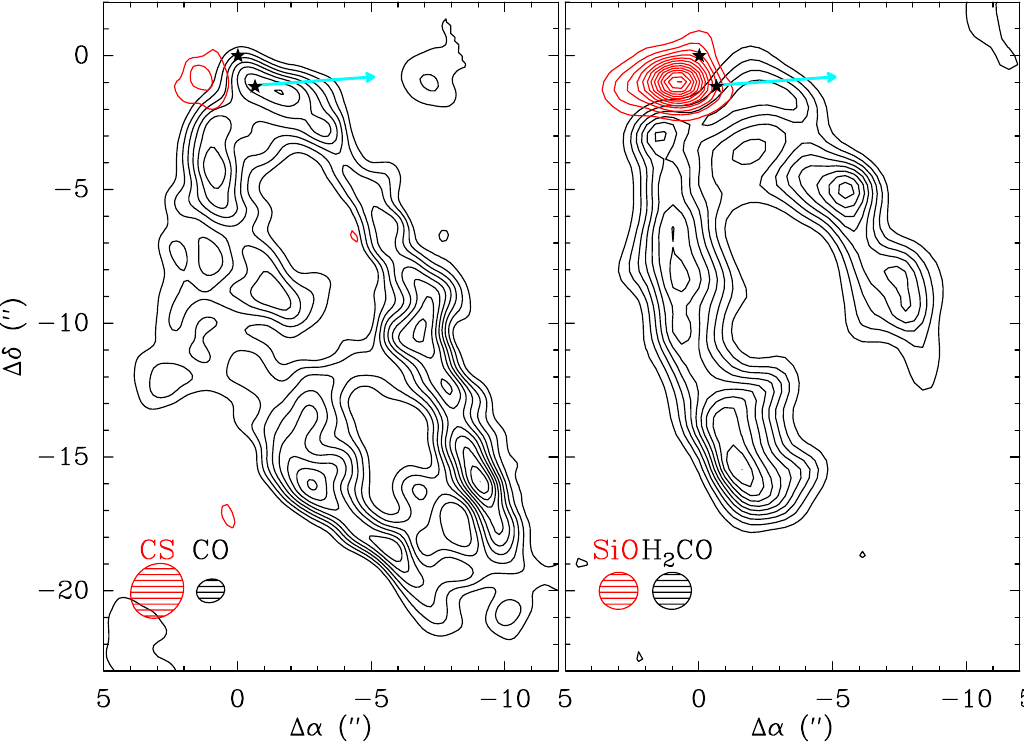}
\caption{\label{fig02} Interaction between the southern low-velocity outflow lobe of A, at $[-8;-6]\kms$ (black contours) and the Eastern lobe of the jet from B (red contours), as observed with the IRAM interferometer in the molecular lines of CO $J$=2-1 and 
CS $J$=2-1 (left), and  SiO $J$=2-1 and H$_2$CO (right).
}
\end{figure}

 Close to the driving protostar A, a spatial overlap is observed in the high-velocity emission of the A and B jets. The blueshifted component of the high-velocity B jet has propagated over a distance of about 5000~au westwards from the protostar. In contrast, the redshifted component seems to have propagated about $~3300$~au eastwards. The spatial distributions of the outflows appear to overlap, at least in projection, in the plane of the sky. { Near source B in Figure~\ref{fig01} (left panel), there is a decrease in CO emission (by 4´´) from the jet at velocities $>50\kms$ (blue contours) , which is not seen for the northern jet. { We suggest that this reduction in CO emission occurs due to an interaction between the primary jet and a second collimated flow, where the pressure gradients are the forces changing the individual outflow dynamics. Observationally, this effect manifests in projection as a decrease in the velocity in the original plane of motion or a smaller projected size for this secondary outflow.} In the right panel of Figure~\ref{fig01}, one can see the flow ejected by source B, and { it could be because} this flow moves in one direction (to the right in this image), but it does not evolve in the same manner in the opposite direction. It can be because of an interaction of the jet (launched by source B) with the medium through which it propagates.}  

 In addition to the overlap of the CO and SiO signatures of the jets and their outflows, \citet{Schutzer2022} showed that the redshifted component of the B jet overlaps with a minimum of emission in the southern low-velocity cavity walls of the A jet. The detection of SiO, CO, CS and H$_2$CO (Figure \ref{fig02}) testifies to the presence of a somewhat "active" shock, and the most likely hypothesis is that the B jet is impacting the southern low-velocity cavity driven by A jet.  Furthermore, there is an 7$^\circ$ misalignment between the northern jet axis and the southern jet axis that can be seen in Figure~\ref{fig01} and \ref{fig02} (left panel) where {we note an asymmetric structure in the cavity }\citep{Schutzer2022}.  Figure~\ref{fig02} also shows the flow of source B with the cyan arrow. This is one of the most convincing cases of jet interaction in a protobinary system.

Then, we estimate the probability for the two outflows to coincide spatially. First, we approximate that both outflows, produced from sources A and B, have a biconic shape. An observer from source B would see two spherical triangles projected in the sky from their point of view with height $L \cos i$ and base $L \cos i \tan \alpha$, where $i$ is the outflow A inclination angle with respect to the plane of the sky and $\alpha$ is the A cone opening angle. Then, assuming source A as the pole of a spherical system and at a distance $r$ from the origin of this system at source, the solid angle covered by a triangular section could be approximated as $\Omega=\phi(1-\cos \theta)$, where $\tan \theta=\frac{L \cos i}{D(1+\sin i)}$ and $\tan \phi= \frac{L\cos i \tan \alpha}{D(1+\sin i)}=\tan \theta \tan \alpha$. Additionally, the outflow produced by B would make a solid angle $\omega=2 \pi(1-\cos \beta)$, where $\beta$ is the opening angle of the B cone. Finally, the probability can be estimated as the ratio of the solid angle compared with the solid angle of half a sphere.
\begin{equation}
P=\frac{\Omega_1+\Omega_2+\omega}{2 \pi}. 
\end{equation}

\begin{figure}
    \centering
    \includegraphics[width=1\linewidth]{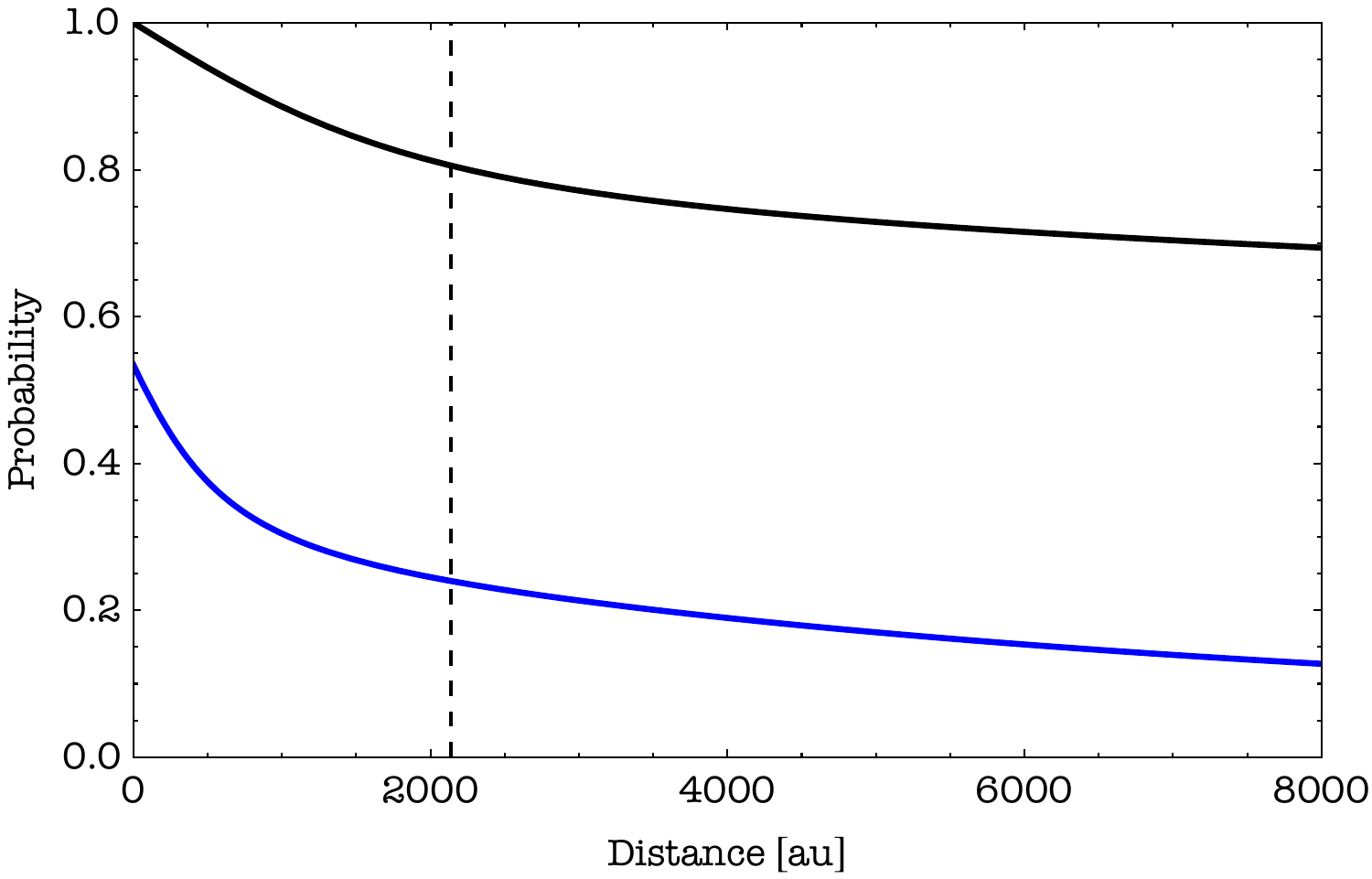}
    \caption{{Probability of two outflows overlapping for: two wide cavities (in black continuous line) and for two collimated jets (blue line). The dashed line represents the observed separation of Cep E-A and Cep E-B, which could be larger because of projection effects.}}
    \label{fig:probability}
\end{figure}

{In Fig. \ref{fig:probability} we estimate the probability of an encounter using an opening angle of $90^\circ$ and $15^\circ$ for the Cep E-A cavity and jet, respectively, and an opening angle of $60^\circ$ and $15^\circ$ for the cavity and jet produced by source B, respectively.
We have used an extension $L=8000$~au, where these opening angles seem reasonable, and the separation of the sources is a free parameter, since they have an apparent separation of 2136 au (dashed line), which could be larger due to projection effects. Then, the lines represent the probability of the encounter between cavities (black) and jets (blue).
This model still understimates the collision probability, which is already high in the case of the cavity-cavity encounter, $P\sim80\%$, since we are not taking into account their three-dimensional structure. Therefore, we expect that the cavities associated with the Cep E binary system are spatially overlapping while their driving jets are not necessarily colliding ($P\sim20\%$).}

\section{Numerical model}
\label{Sec:3}

To describe the { collision between}  { the envelopes and the main flow, jet)} and characterize the observational signatures produced by this interaction, we have selected Cep \,E to obtain a significant number of physical constraints required to produce a consistent model. Cep \,E is a protobinary system, where each component produces a high-velocity outflow \citet{Ospina-Zamudio2018}. As discussed before, there is evidence of interaction between the powerful outflow produced by the component Cep\,E-A and the jet produced by the component Cep\,E-B. Also, we considered that each material injection zone is embedded in a density stratification associated with a protostellar core, as modelled by \citet{Crimier2010}.  Then, we modelled the { collision} of the outflows produced by two high velocity jets moving in this stratified environment. This section describes the numerical model, first explaining the code used to perform our simulations. Then, we give the physical constraints obtained from the Cep\,E protostellar source.

\subsection{Numerical setup}

The presented numerical simulations (Table \ref{tab:models}) are carried  grid inside a physical domain $\left[x,y,z\right]\, \in\,\left[\left(0,1\right),\left(0,4\right),\left(0,1 \right) \right]\times 10^4 {\rm au}$ with a spatial resolution of $4.88\, {\rm au}$ in each direction. Then, the numerical box is a rectangular prism and, considering that the origin of the Cartesian plane lies at the center of a square face of the box, with coordinates $\left[0,0,0\right]$, we inject 
high-velocity gas at two positions into the grid to reproduce the propagation of two protostellar jets, similar to those observed in Cep\,E.
The first jet is the main jet: it is launched from position  $\left[0,0,0\right]$, and the other injection zone is the secondary jet, injected at position $\left[1500,1500,250\right]$ ${\rm au}$.

We performed our  3D hydrodynamic simulations using the {\sc Guacho} code\footnote{The {\sc Guacho} code is a free access code maintained on http://github.com/esquivas/guacho}  \citep{Esquivel2009}. This code solves the near-conservation laws governing the gas dynamics and the neutral hydrogen rate in a Cartesian grid.  The complete set of equations is:

\begin{eqnarray}
 \frac{\partial \rho}{\partial t}+\vec{\nabla} \cdot \left(\rho\,\vec{v}\right)&=&0,\label{eq:gas:density}\\
 \frac{\partial \rho v}{\partial t}+\vec{\nabla} \cdot \left(\rho \,\vec{v}\cdot \vec{v}+p\right)&=&0,\\
\frac{\partial e }{\partial t} +\vec{\nabla}\cdot \left[\vec{v} \left(e+p\right) \right] &=&G_{\rm rad}-L_{\rm rad}\,,\label{eq:ener}
\end{eqnarray}

\noindent where, $\rho$, $\vec{v}$, $p$ and $e$ are the density, the velocity, the thermal pressure, and the total energy density of the gas, respectively.  $G_{\rm rad}$ and $L_{\rm rad}$ are the
energy gains and losses due to radiation. We assume a standard ideal gas law for the closure of 
the gas total  energy density and
the gas thermal pressure is related by $e=\rho \vec{v}^2/2+P/(\gamma-1)$, where P is the total pressure, and the ratio between specific capacities is $\gamma=1.4$. The code is described in detail in \citet{Rodriguez-Gonzalez2019}. Energy is lost via radiative cooling, which is included as described by \citet{Raga-Reipurth2004} for the atomic gas, temperatures about thousands of kelvins. For lower temperatures, we have included the parametric molecular cooling function presented in \citet{Kosinski-Hanasz2007} and \citet{Rivera-Ortiz2019}. We have also considered heating resulting from the hydrogen's cosmic ray ionization rate, proposed in \citet{HETAL09}. { The two protostars are separated by a distance of \(2136~\mathrm{AU}\), which makes the Cep E outflow age to be a fraction of the orbital period of the system, and the interaction between their supersonic jets occurs at large distances from both protostars, where the escape velocity is much lower than the outflow velocities. With this configuration, gravity and self-gravity are not considered in the models investigated in this paper. Additionally, in low mass star forming regions, magnetic pressure is significantly weaker than thermal and ram pressure in the supersonic outflow at a distance more than thousands of au from the protostellar disk, since a large faction of magnetic energy is already converted to kinetic energy and the magnetic field is in the order of $10^{-5}$-$10^{-3}$ G for similar jets in the pre and post shock gas, respectively, in other HH outflows. \citep{Pudritz19}. Thus, magnetic fields are not a dominant factor in the asymmetries observed in the Cep E outflow. 
The effects of gravity, magnetic fields and even the radiation on the accretion disk and outflow launching have been extensively studied in previous works (e.g., \citealt{2006A&A...460....1M, Mignon-Risse2021a, Mignon-Risse2021b, OlivaAndKuiper2023}). 
}

%


\subsection{{Physical constraints}}
\begin{figure}
    \centering
    \includegraphics[width=0.7\columnwidth,angle=0]{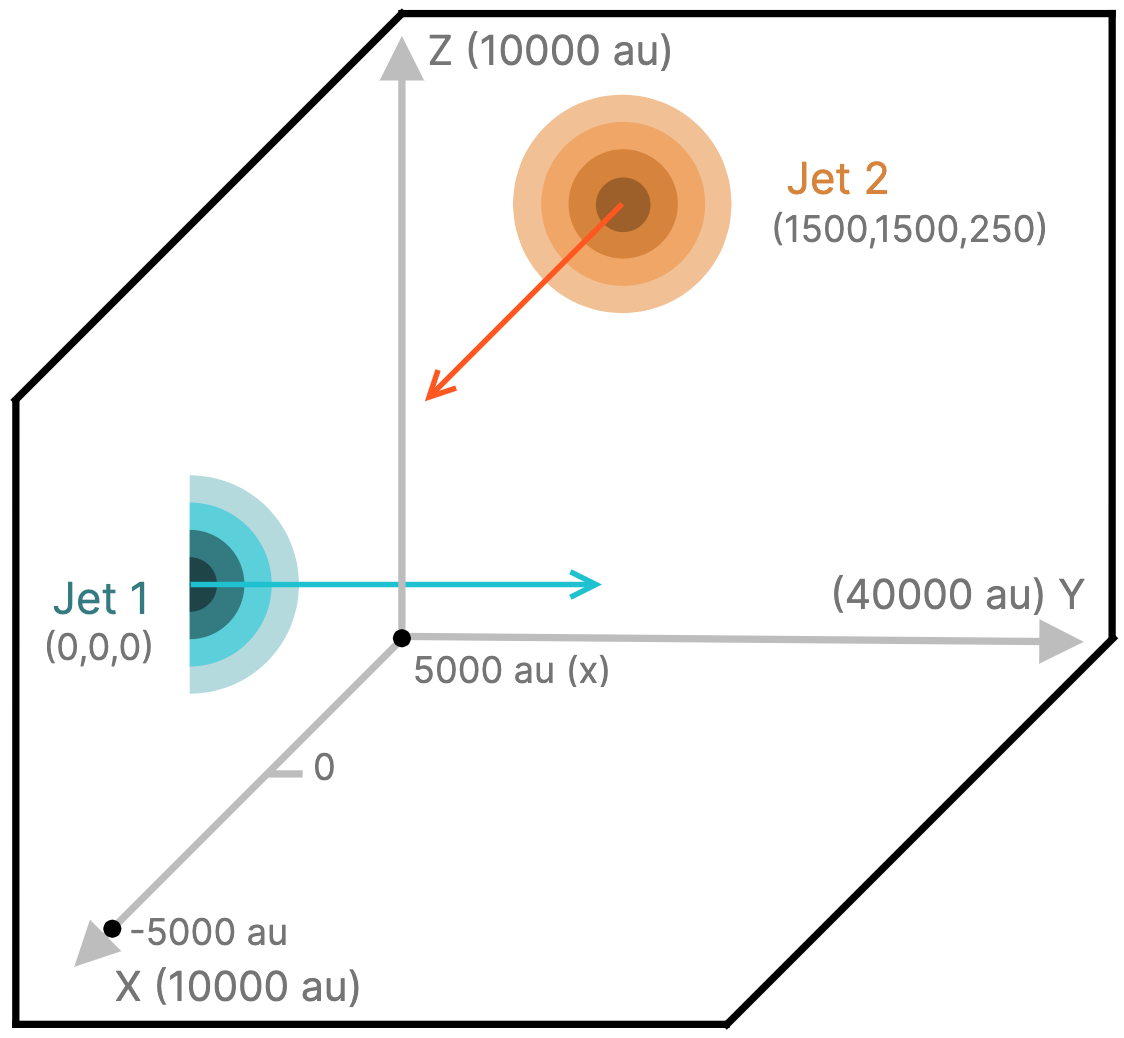}
    \caption{Two protostellar sources inject high velocity material. The gas density around each source is stratified with a power law. { In our simulations the jets injections are not meant to collide if they preserve their width, but, since they expand thermically and by entrainment, the evolved jets, that is, high velocity material, could collide.}}
    \label{fig:model}
\end{figure}

In Figure \ref{fig:model}, we have decided to use one jet (the primary jet) that is propagating in the {\it y} direction and a secondary jet that propagates in the $x$ direction. We have decided to avoid a direct collision of the two jets, since the direct collision has a lower probability of occurrence, { but, as they evolve they could still interact.} Both propagation directions are perpendicular in the $xy$ plane, but they have an impact parameter of 250 au in the $zy$ plane, 
 so that the material injected by the secondary jet comes out of the primary cavity with more momentum and has a strong impact on this cavity.

\begin{table}[!h]
\label{tab:doi}
\begin{center}
\caption{Characteristics and parameters of the simulations.}
\begin{tabular}{c  c  c }
\hline \hline
Model       &
\multicolumn{1}{c}{$t_{s}$}  &  {$v_{j,s}$}\\
           & [yr]    &  [km~s$^{-1}$]\\ 
\hline
M1        & ---  & ---  \\ 
M2        & 660.  & 100.      \\
\hline\hline
\end{tabular}
\label{tab:models}
\end{center}
\end{table}

\begin{figure*}
    \centering
         \includegraphics[width=1.7\columnwidth]{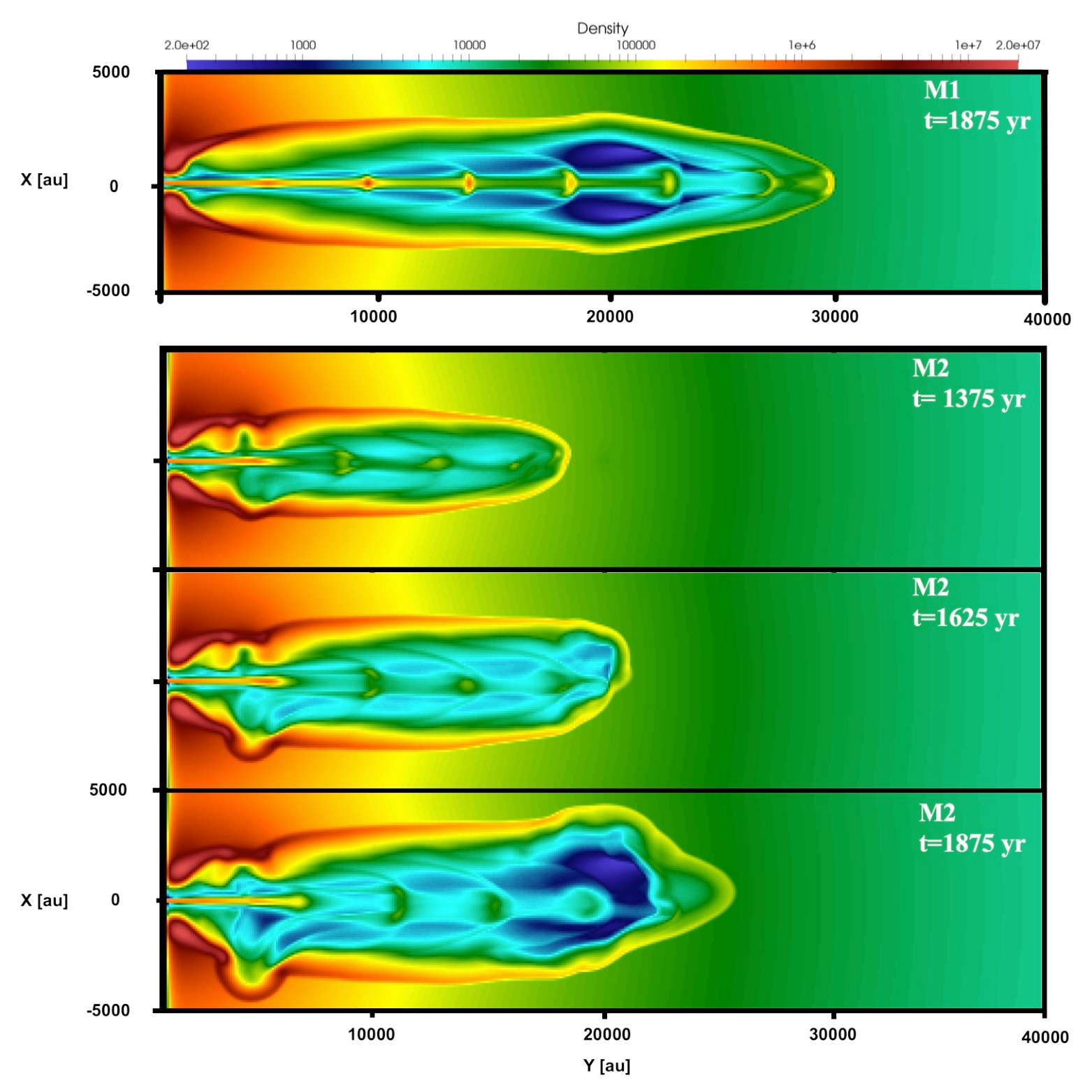} 
\caption{(Upper panel) Snapshot of the gas density of a single jet (model M1) at $t=1875$ yr. One can see {the work surface structure can be seen due to the consideration of the variable jet}. The envelope shows a higher density at the "base" of the jet and a lower one towards the front. 
(Lower panels). Snapshots of the gas density of the jet interaction model (model M2) at 3 evolutionary times, t=1375, 1625 and 1875 yr (top, middle and bottom panels). Considering the estimated age of the Cep E protostellar outflow, $\sim$1500 yr \citep{Schutzer2022}, it shows different stages of evolution around this age. All the snapshots are present on the $xy$ plane.
}
    \label{fig:M1}
\end{figure*}

Both of the jets are considered a cylinder with radius of r$_J$=$63\, {\rm au}$, and an injection length of $r_L$=$42\, {\rm au}$, and both jets have the same thermodynamic properties, that is, a numerical density of $n_j=1\times 10^6$~cm$^{-3}$ and a temperature of $T_j=250$~K.  The primary jet was injected using a time velocity variation, as was proposed in \citet{Rivera-Ortiz2023}:

\begin{equation}
  V_j=V_{j,m} [1+\delta v \cos(2\pi t/\tau)]  
  \label{eq:period}
\end{equation}

where, $V_{j,m}$ is the mean jet injection velocity = $200 \,$km~s$^{-1}$,  relative amplitude variability $\delta v_j =0.08$, which implies velocity variations of $16$~km~s$^{-1}$, and $\tau=130$ yr, which is twice the period determined by \citet{Schutzer2022}.  The mass injection can be calculated using \citep{Rivera-Ortiz2023}:
\begin{equation}
    \frac{\dot{M}}{ 10^{-6}\rm{M_\odot yr^{-1}}}=1.5\left[\frac{r_j}{\rm{50~au}} \right]^2 \left[\frac{n_j}{\rm{10^6 cm^{-3}}}\right] \left[\frac{v_j}{\rm{165 km s^{-1}}}\right] = 2.88
    \label{eq:mdotjet}
\end{equation}

The secondary jet is turned on in $t_{s}=700$ yr after the primary jet was initially ejected. This timing was determined based on the estimated age of the flow and the location of material dispersion in the southeastern low-velocity cavity, which is located about a half distance considering the jet length and a dynamical age of 1500 yr \citep{Schutzer2022,Rivera-Ortiz2023}.
We have changed this value to inject the secondary jet at different distances from the main jet wall cavity.  The velocity of the secondary jet is $V_{j,s}=100$ km~s$^{-1}$. {We define}
\begin{eqnarray*}
r&= & [x^2+y^2+z^2]^{1/2}\\
r_b&= & [(x-x_b)^2+(y-y_b)^2+(z-z_b)^2]^{1/2}\,,
\end{eqnarray*}
{as} the radial distances to the primary source, $r$, and to the secondary source, $r_b$. 
Then, considering both envelopes, the initial gas density and temperature distributions are imposed as follows
\begin{equation}
    n(r)=
     \frac{n_0}{1+(r/r_0)^{\alpha_1}} +\frac{n_0}{1+(r_b/r_0)^{\alpha_1}}\,,
     \label{dens_dist}
\end{equation}
{ which is the profile proposed by \citet{Crimier2010}. Considering that each term on the right-hand side of this equation represents the density profile corresponding to each source, with a parameter \(\alpha_1 = 1.9\), this profile corresponds to a gas mass of \(0.3~\mathrm{M_\odot}\) up to \(500~\mathrm{au}\) from the central source and a gas mass of \(0.14~\mathrm{M_\odot}\) up to \(300~\mathrm{au}\) from the secondary source. This value is in reasonable agreement with the  mass reported by \citet{Ospina-Zamudio2019} and the mass-luminosity relationship for young stellar objects \citep{Motte2001}.}


The region also has a temperature profile given by

\begin{equation}
    T(r)=\frac{1}{2}\left[
     \frac{T_a}{1+(r/r_0)^{\alpha_2}} +\frac{T_b}{1+(r_b/r_0)^{\alpha_2}}\right]\,,
\end{equation}
where $T_a=T_b=300$ K is the central temperature of each core and $\alpha_2$ is $-0.8$ in the range between $50-300$ K and $-0.4$ in the range $7-50$ K, according to \citet{Crimier2010}.

\section{Results}
\label{Sec:4}

We have carried out two hydrodynamical simulations to analyze the effect of the { collisions} of the two jets { in the dynamics and morphological evolution of the gas injected by the protostellar objects}. The first (M1) has a single jet where the material is injected with a velocity variability, described by Eq.~\ref{eq:period}, into a stratified interstellar medium, allowing its dynamic evolution. We have allowed the material to evolve for 1875 yr. That is, there are more than 14 periods of maximum velocity in the injection of the jet. In the second model (M2), we have considered the evolution of a single jet for 660~yr, and we have turned on the second jet, which propagates perpendicularly regarding the first of them. We have followed the gas evolution since the secondary jet reached the limit of the simulation box (at 1875 yr, as well).

\subsection{Morphology evolution}
The morphology of the single jet (model M1) evolution is presented in Figure \ref{fig:M1} (Top). There is a snapshot of the central slice of the density distribution of the M1 model, where only the evolution of a single jet has been considered. As expected, in the density, the working surfaces resulting from the velocity variation imposed on the jet's injection zone and the material environment's envelope were shocked by shock waves expanding laterally with subsonic velocities with low Mach numbers. {In the region closest to the material injection point, a very dense envelope is visible}. From this figure, {one} can see how the density on the outer wall is greater at the base of the jet; this is because there is a much denser environment in this region, and we have 7 knots or visible work surfaces that structure the gas in the cavity, from the centre. For this model, {one} can see that the envelope reaches a width of 6000 au, when the length is approximately 30000 au.

\begin{figure*}
    \centering
\includegraphics[width=1.7\columnwidth,angle=0]{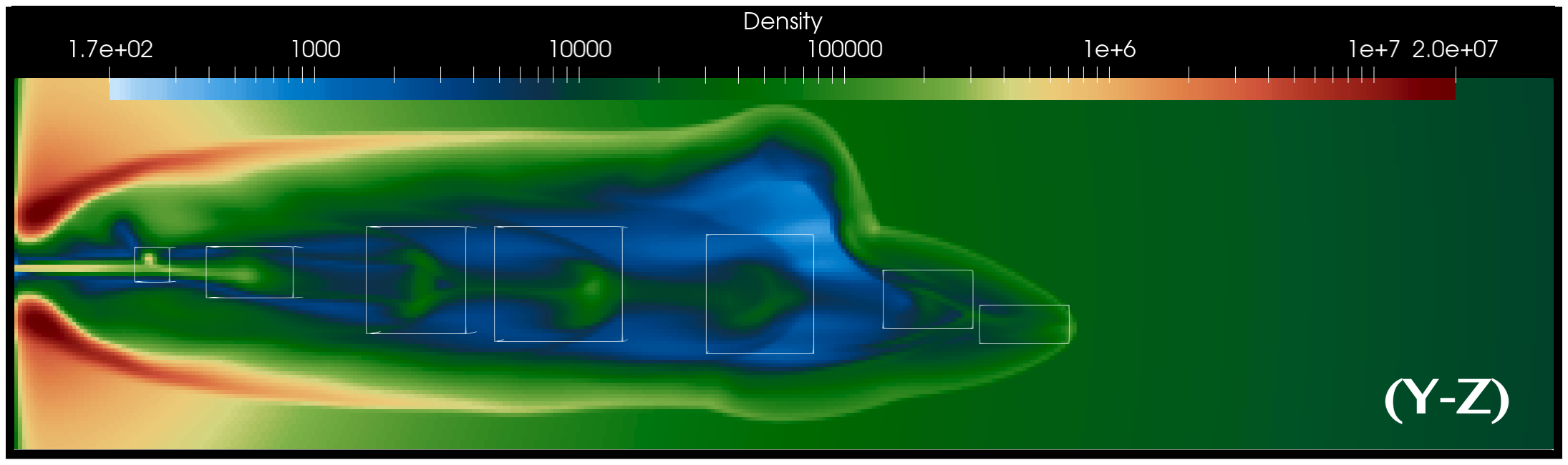}
    \caption{Snapshots of the gas density of { two} jet { collisional} model (model M2) in the plane Y-Z. The colour represents the density, and the white boxes are the position of the knots. Using the boxes we can find the mass of the knots.}
    \label{fig:M2YZ}
\end{figure*}

\begin{figure}
    \centering
    \includegraphics[width=0.9\columnwidth,trim=0 0 0 1cm,clip=true]{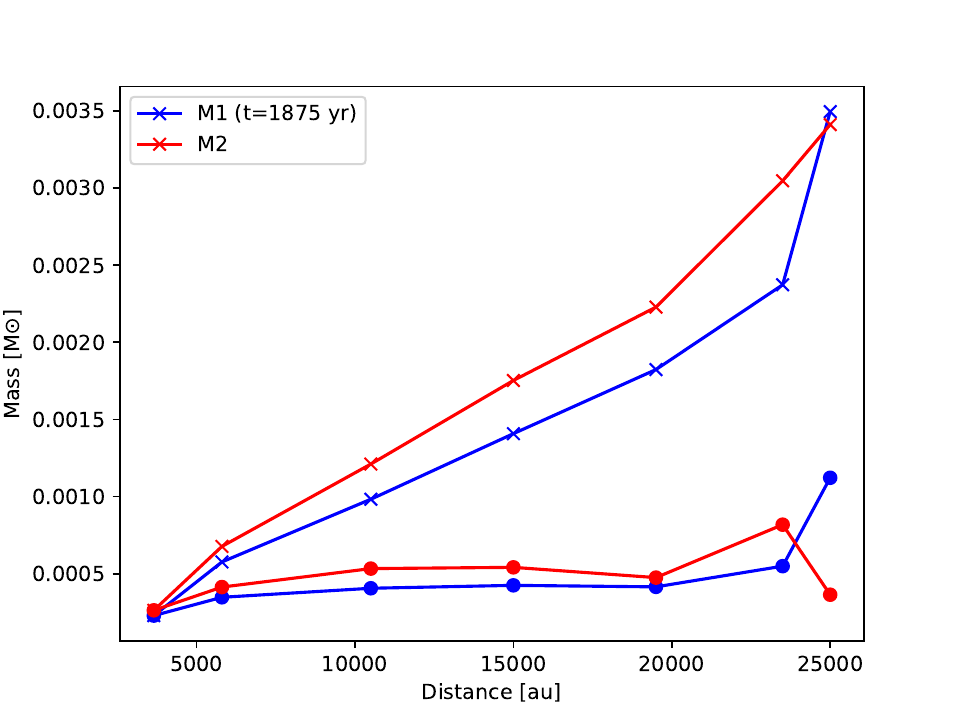}
    \caption{The mass of the knots of models M1 (in blue circles) and model M2 (in red circles) and the accumulated mass are shown in crosses. The figures show that the mass in the knots of the model with { two jet collisions} is systematically greater for all knots than those of the M1 model, except for the furthest knot, which is difficult to determine.}
    \label{fig:mass_dist}
\end{figure}

We have also run a simulation considering the mass and energy input rates of two different jets. We are interested in evaluating the effect of a secondary jet (which we have considered continuous) on the evolution of a variable jet that we consider the primary one. We have added the secondary jet, assuming that it originates within the envelope formed by the supersonic movement of the envelope towards the interstellar medium, where it propagates. Figure \ref{fig:M1} shows the central slice density evolution of the gas in model M2 (color maps), with two winds colliding. The upper, middle and bottom panels present the evolutionary time 1375, 1625 and 1875 yr, and one can see the secondary jet position in [(3500), (5250)] au in this image. {One} can see the evolution of the gas injected by the secondary jet at 1375 yr (top panel) in the disturbance of the main envelope where the gas from the secondary jet impacts the wall (bottom wall in this figure). Moreover, at 1375 yr, the frontal part of the envelope does not seem to have significant changes; however, as the interaction continues, the envelope deforms, and this perturbation grows, creating a considerable asymmetry on the same side where the secondary jet is injected. The cavity reaches a width of about 8500 au, at 1875 yr, this being about 30\% wider than the cavity formed in single jet evolution. 

Note that the disturbance in the primary jet envelope is related to the first moment the two jets interact, a relic of the { collision}. The interaction between the material coming out of the secondary jet that impacts and intervenes in the flow of gas that makes up the primary jet changes the general direction of the flow.
The size of this region of impact increases, forming a bubble; the evolutionary times 1625 and 1875 yr are presented in the middle and bottom panels, respectively. At 1875 yr (bottom panel), a bubble of about 2000 au is formed, containing material injected by the primary and secondary jet. The densest part of the shell of the bubble is formed because of the jet and original cloud gas supersonic collision. Therefore, the gas envelope formed by the thermal expansion of the envelope has a faster-evolving region due to the kinetic energy injected by the secondary jet into this cavity.
The primary jet outflow suffers a deviation in its trajectory of approximately 5 degrees. The deviation of the flow has its most significant effect in the interaction with a part of the cavity that moves at much slower velocities than the front part of the shell. In this region of impact,  the envelope has a lateral movement of a few km~s$^{-1}$. The { collision} between the main jet and the inner part of the envelope has an essential effect on the overall evolution of the bubble. The surface where the jet hits partially reflects it, making its movement conditioned by the shape of the wall where it slides.

The cavity formed by a single jet appears to have a lower density because much of the injected material remains on the working surfaces, {where two shocks are formed, one toward the source and the other away from the source}, created by the variability of the jet. The lateral growth induced by the secondary jet can be observed as jet envelopes much larger than those expected in the single jet models. The jet beam also appears almost three times wider, and the knots or working surfaces are broader than the primary jet flow. In the scenario with two jets, the working surfaces appear much less dense or compact and, in some cases, seem to have dissipated (see Figure~\ref{fig:M1} bottom panel), creating a region at the front of the cavity that appears to expand due to internal pressure.

\subsection{Forces analysis}

{ To obtain the effect of the interaction between molecular outflows and to compare it with other forces that may play an important role in the evolution of young outflows, we computed the material derivative of the velocity matrices under the assumption of a stationary state. This procedure allows us to obtain acceleration matrices along each axis, from which the force per unit volume can be derived by multiplying the accelerations by the number density.}

\begin{figure*}
    \centering
    \includegraphics[width=0.8\linewidth]{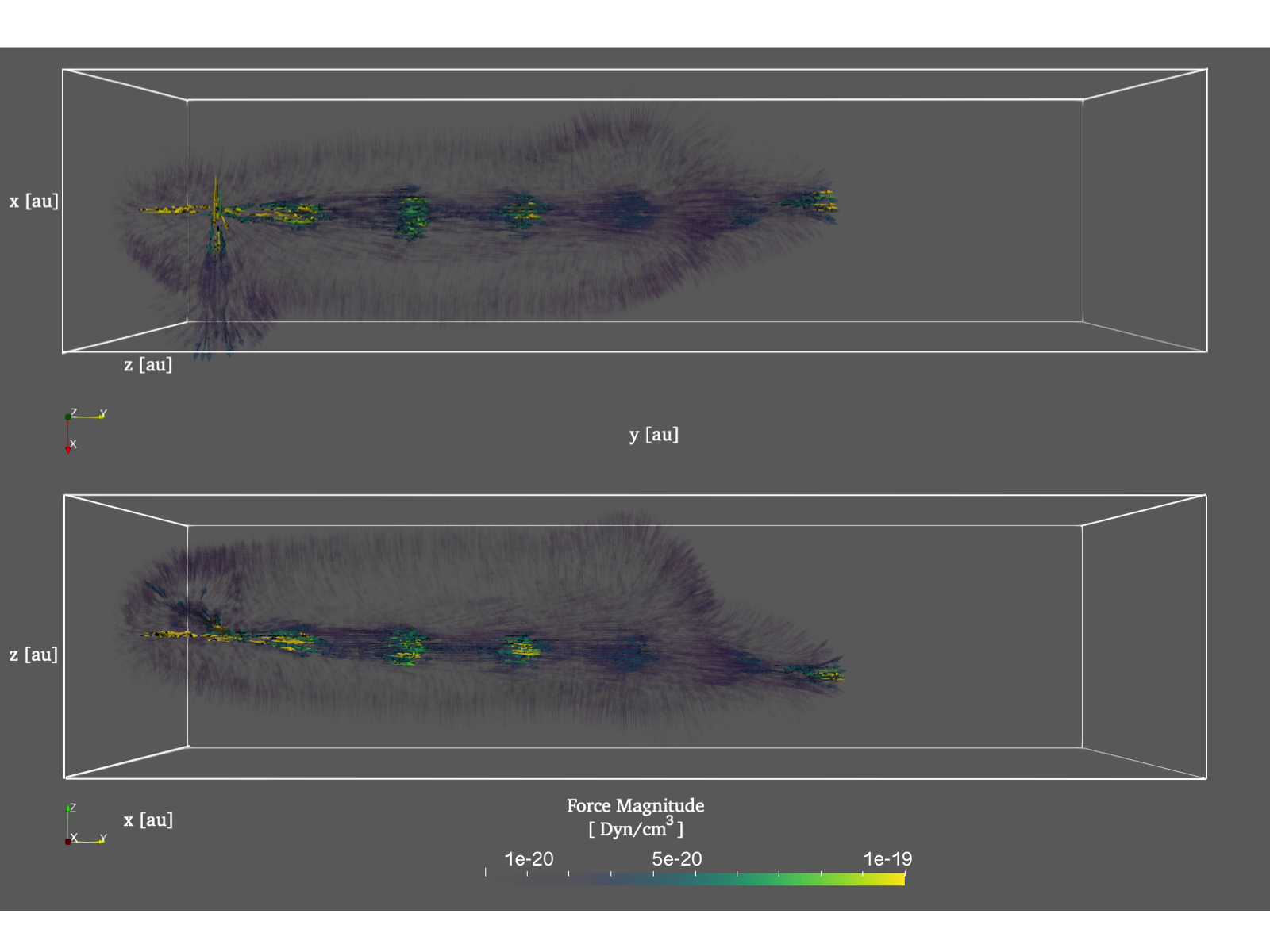}
    \caption{The resulting force field, where arrows indicate the three-dimensional direction of the force and a colour scale encodes its magnitude.The upper panel displays the y–x plane, while the lower panel shows the y–z plane.}
    \label{fig:forces}
\end{figure*}

{ Figure~\ref{fig:forces} presents the resulting force field, where arrows indicate the three-dimensional direction of the force and a colour scale encodes its magnitude.  In both panels, the figure illustrates that the force due to the gas flow is strongest in the jet injection region, at the knots or working surfaces formed by the temporal variability of the primary jet velocity, and at the leading edge where the jet encounters the ambient gas cloud. Both the upper and lower panels also reveal the expansion of the envelope, albeit with lower forces. Notably, this expansion is more pronounced up to two knots behind the leading knot, highlighting that the future evolution of this object will be highly asymmetric in its frontal region.}

{ From the y–x plane (upper panel), the interaction between the flows further reveals: (a) a change in the force of the secondary flow along the direction of propagation of the primary jet, (b) the formation of a protrusion or secondary bubble generated by the momentum injected by the secondary source, and (c) broadening forces that affect the evolution of the primary jet. From the y–z plane (lower panel), one can observe that the secondary flow develops an expansion triggered both by its interaction with the primary jet and by the resulting force from the combined flows.}

{ To quantify the effect of the magnetic field with respect to the supersonic motion of the two flows present in Cep E, we can calculate the magnetic pressure $P_B$, withe the ram pressure $P_{ram}$, one can be calculated $P_B=\frac{B^2}{4\pi}$, where $B$ is the magnetic field, while the ram pressure is $P_{ram}=\mu m_h n v^2$. The force produced by each pressure is $f=\nabla  P$, which we approximate to $f=\frac{P}{\Delta x}$, assuming the pressure changes in a scale $\Delta x$, that is, the cell size in our simulations. For values of $B=1$~mG, $v=100$~km~s$^{-1}$, $ n \sim 10^{6}$~cm$^{-3}$, the ratio $\frac{f_{ram}}{f_{B}}\sim 10^{-4}$, which helps us to scale these forces. For lower velocities, around a few km~s$^{-1}$, both forces are comparable, where $B$ is as large as $1$~mG, but typical values away from the protostellar core, $10^{-6}-10^{-4}$~G \citep{Pudritz2019}, make the magnetic force very small compared with the ram pressure. This value for $B$ is derived from the analysis by \cite{Lefloch2015}, who performed a grid of models to reproduce the CO emission in the Cep~E jet shocks.}

\subsection{Jet deviation and pseudo-precesion}

The { collision} between the { outflows} appears different when observed from another reference axis. For instance, in the $yz$ plane (Figure~\ref{fig:M2YZ}) { shows the envolved interaction between the outflows. In the region of the first knot close to the source, the secondary jet shows like a point up the principal jet.} On the other hand, the flow diversion caused by the {interaction} between the { outflows} is {noticeable because the nodes move away from the center of the box where the injection is generated, creating the impression of descending}. 

\begin{figure}[h]
    \centering
    \includegraphics[width=\columnwidth,angle=0]{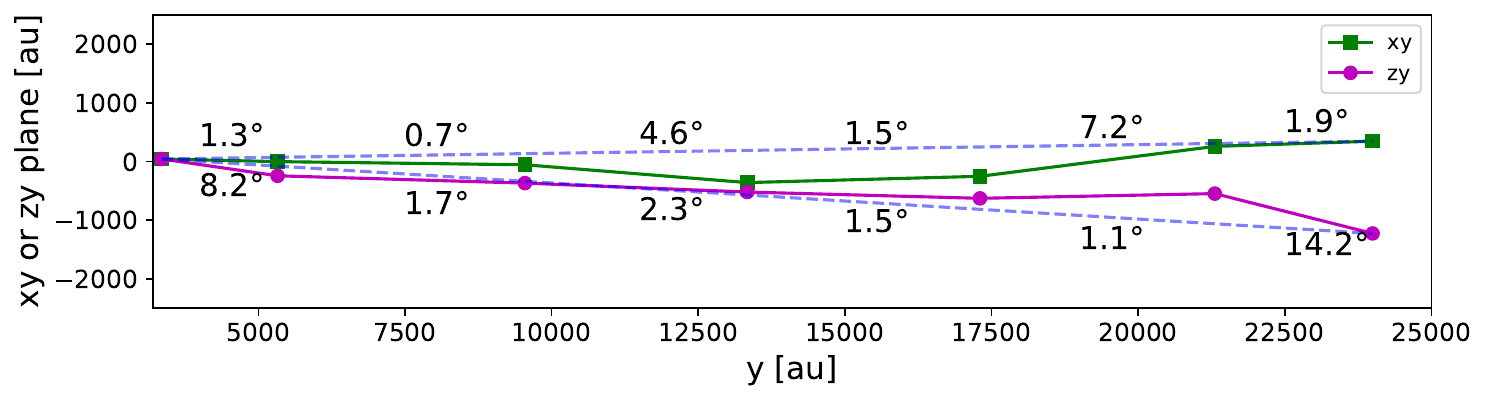}
    \caption{The position of the knots respects the $xy$ (green/square) and $zy$ (pink/circle) planes. The dashed line represents the angle between the first and the last knot. For the $xy$ plane, an angle of 0.84 degrees is observed, and for the $zy$ plane, an angle of 3.51 degrees. Furthermore, the angles for the knots are presented. These angles are calculated using two consecutive knots, representing the inclination of the line drawn between the two knots.}
    \label{fig:pseudo-prec}
\end{figure}

{  On the other hand, to characterize the deviation of the main flow}, we have placed some boxes centered on the barycenter of each knot produced by the work surfaces. As seen in this plane, the $yz$ plane, one can observe how the main flow is deflected due to the moment injected into the secondary flow. The secondary jet was injected at 250~au over the plane of the primary jet.  Therefore, the main flow deviated in the $z$ direction, producing an asymmetric envelope evolution. This asymmetry in the evolution of the envelope is driven by the flows of the main jet, which interact with the inner part of the envelope, thereby changing the flow direction. Furthermore, one can evaluate the asymmetric evolution of the nodes and observe variations among the different planes. Figure~\ref{fig:pseudo-prec} shows the position of the knots, green and pink points in the $yx$ and $yz$ planes, respectively, of the M2 simulation. The $yz$ plane shows the global deviation of the jet is 3.5 degrees in the opposite direction to the injection point of the secondary jet. However, in the $xy$ plane, the deviation concerning the $x$ axis is negative for the first five nodes (away from the direction of the secondary source) and positive for the last two. This false precession is caused by the interaction between the jet and the internal wall of the cavity on the opposite side of the injection due to the secondary jet. The maximum deflection due to this process is about 2.3 degrees in the opposite direction to the secondary jet and 5.11 degrees in the other direction. In addition to the model M2 jet being wider than the model M1 jet by approximately 15\% (see Figure \ref{fig:M1}), it will be observed with a precession effect that makes it appear much more expansive than in the case of single jet evolution.

{At $t = 1875$~yr, the total mass injected by the primary jet in models M1 and M2 is $\sim$0.005~M$_\odot$ (see Eq.~\ref{eq:mdotjet}). The redistribution of this material within the computational domain is illustrated in Fig.~\ref{fig:mass_dist}. More than half of the injected mass is concentrated in the working surfaces (blue crosses), while a dense bow shock forms at the head of the cavity, corresponding to the swept-up ambient medium that receives most of the jet momentum.
The mass distribution along the jet axis shows that the internal knots have comparable masses, typically $\sim4\times10^{-4}$~M$_\odot$, and together they account for about 70$\%$ of the total injected mass. Approximately 10$\%$ of the jet material is found in the surrounding cavity, which has been carved out by the jet and filled with material originating from the jet itself. This matter has been expelled laterally from the beam due to the action of internal shocks propagating within the jet. The remaining $\sim$20$\%$ of the mass is distributed along the jet beam between the knots.}

This contains material from the environment swept up during the entire evolution of the jet. However, the total injected mass (cumulative mass) does not agree with the injected one. Likely, a large amount of injected material, about 30\% in M1 model, is not in the knots but in the cavity dug by the jet.  The last knot, the one furthest from the injection zone,  is much more massive than the internal one because a large amount of the environment has been swept away by the bow shock formed by the supersonic evolution produced by this jet.

\begin{figure*}
    \centering
    \includegraphics[width=1.4\columnwidth, angle=-90]{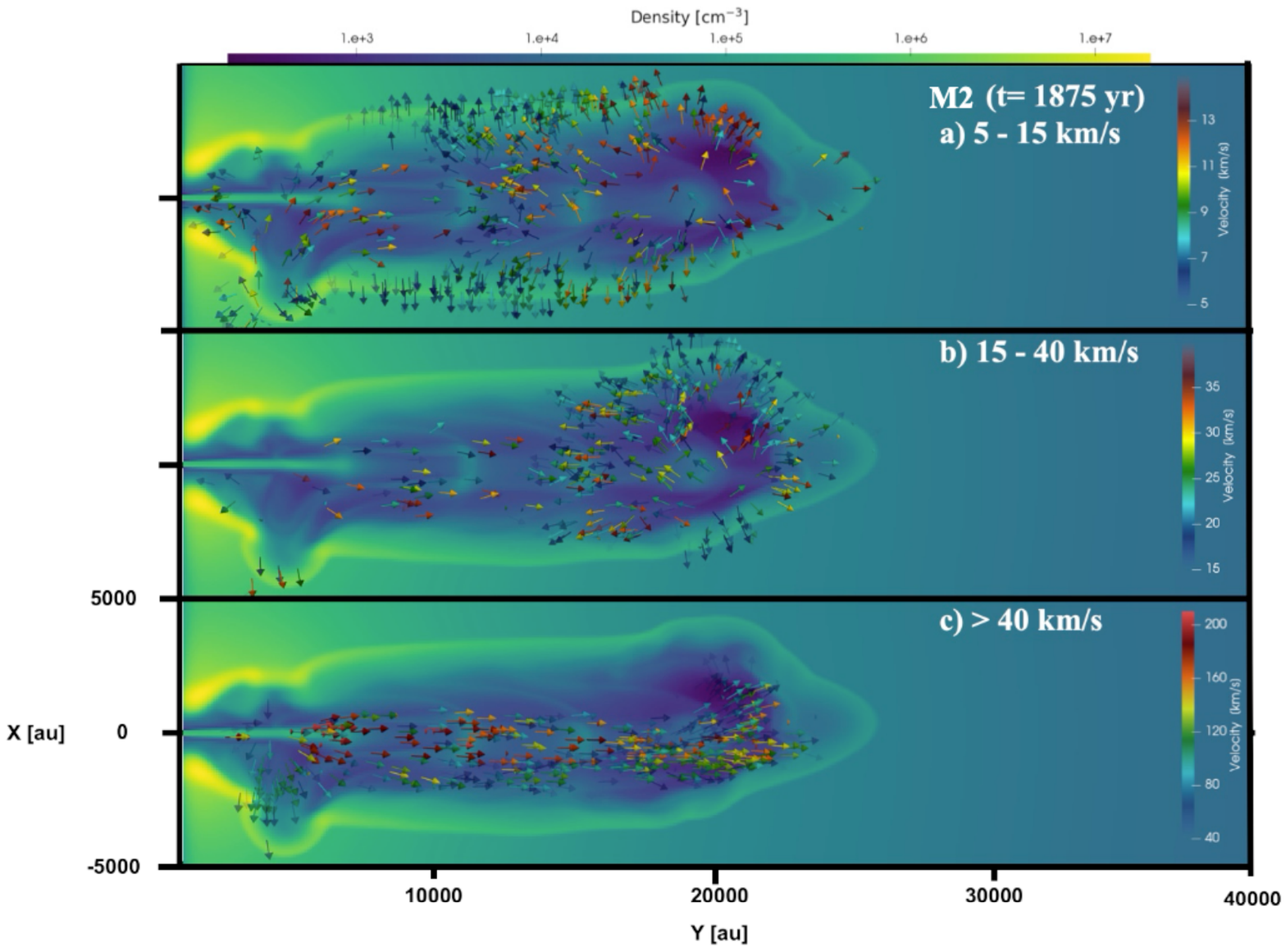}
    \caption{Velocity field for low (from 5 to 15 km~s$^{-1}$), intermediate (from 15 to 40 km~s$^{-1}$) and high velocities ($>$ 40 km~s$^{-1}$), upper, middle and bottom panels respectively.}
    \label{fig:velfield}
\end{figure*}

\subsection{Gas Dynamics}
To analyze in detail the motion of the gas within the envelope, we have {plotted} velocity fields to the density maps for the numerical simulation at 1875 yr; see Figure \ref{fig:velfield}. However, to analyze in detail the dynamic evolution of this gas, we have divided maps with velocity field into 3 velocity channels (in Figure \ref{fig:velfield}): a) 5-15 km~s$^{-1}$ (top panel), b) 15-40 km~s$^{-1}$ and c) $>$ 40 km~s$^{-1}$. The lower velocity channel shows mainly the lateral expansion of the envelope, and the higher velocity channel shows mostly the motion of the gas injected by the jets. Therefore, the channel of velocities between 15 and 40 km~s$^{-1}$ shows the gas dynamics inside the cavity.

In Figure \ref{fig:velfield}  a), in the lower velocity channel, it is observed at  $<$ 7 km~s$^{-1}$ how the envelope expands laterally, moving against the surrounding environment. In the frontal cavity, and in particular, where the largest disturbance of the envelope is found, it can be seen that the expansion grows quasi-spherically with a velocity greater than 10 km~s$^{-1}$, and the gas with higher temperatures, in this velocity range, is pushing the cavity.  For the high-velocity gas, see Figure \ref{fig:velfield}, the jet material moving with a high Mach number propagates in the jet-flow line, but the collision
 with the secondary jet curves, the propagation of the main flow. This deviation causes the main flow to interact with the inner wall of the envelope, deflecting the jet flow due to its collision with the wall. {Changing the direction of the main flow} produces an asymmetry in the envelope that moves in the direction of the injection region opposite to the injection of the secondary flow and is a relic of the initial interaction of the two flows. At the rear of this flow, one can observe the creation of a small bubble caused by the interaction between the secondary flow and the inner wall of the envelope. This {protuberance} reduces the density of the wall, creating a hole in the envelope. It is more visible, particularly in the channel of velocities between 15 and 40 km~s$^{-1}$, where you can see the supersonic expansion of the frontal cavity. This expansion appears spherical, producing a significant increase in the frontal bubble, which produces the great asymmetry observed. {  In Figure~\ref{fig01}, a decrease in CO emission is also observed near source B. One can observe similar behaviour in Figure~\ref{fig:velfield} c) where near the source, no material is observed at higher speeds ($>$ 40 km~s$^{-1}$), which leads to this effect being due to the collision.}



\subsection{Observational predictions}

\begin{figure*}[h]
    \centering
\includegraphics[width=1.9\columnwidth,trim=0 2cm 0 2cm, clip=true]{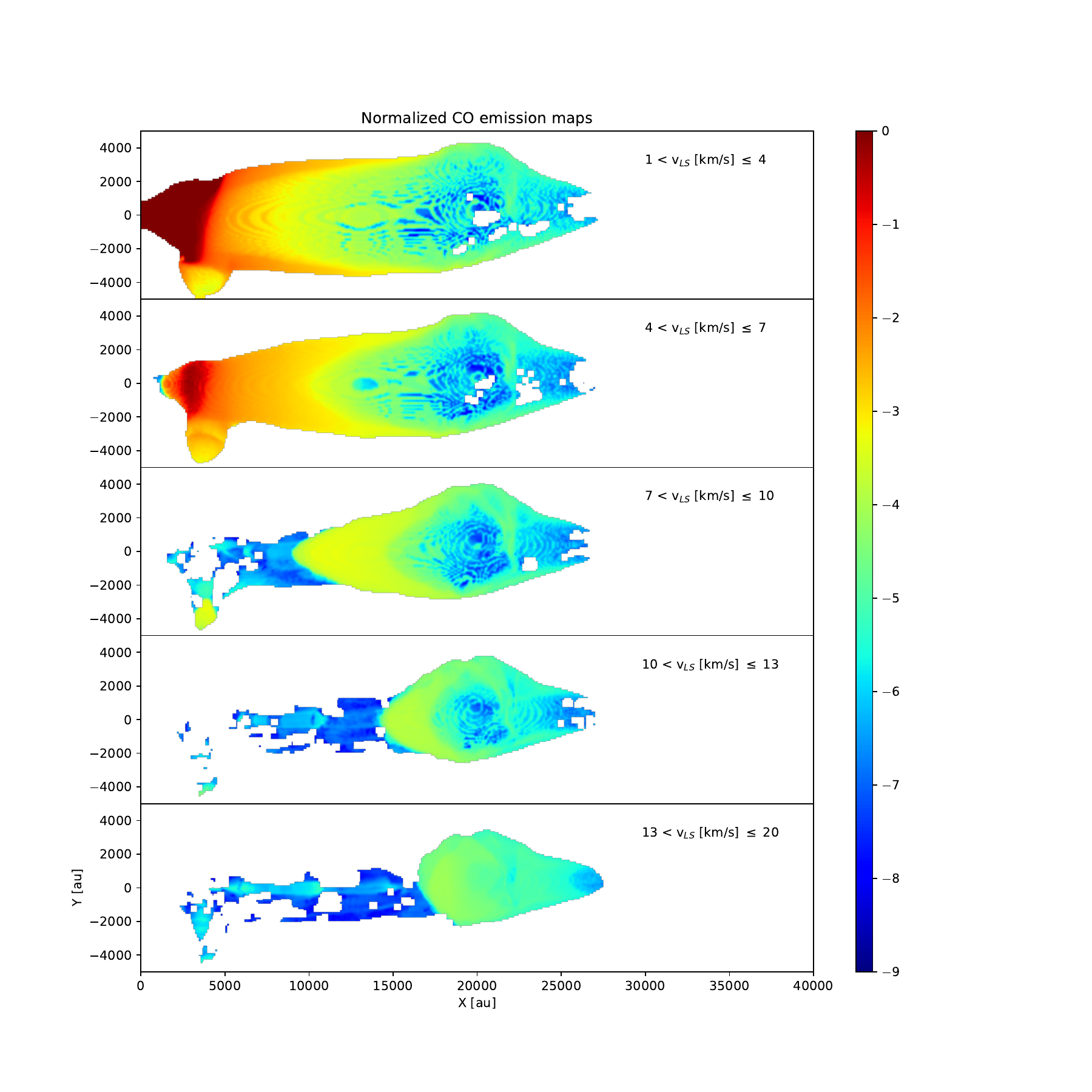}
    \caption{Emission of the M2 model considering channels of velocities. For these maps, the velocities in the direction of the observer are considered. The upper panel shows the emission for the velocities between 1 and 4 km~s$^{-1}$. The emission delineates the outer cavity walls, including the bubble generated by the secondary jet in the bottom left corner. On the other hand, a second bubble can be noticed by the lack of emission at a distance of 2 $\times$ 10$^4$ au. The second panel, 4 and 7 km~s$^{-1}$, shows the inner cavity walls and the leading shock, like the secondary bubble, shows no emission in this velocity range. The third panel, 7 and 10 km~s$^{-1}$, allows us to observe more emission inside both bubbles, and near the base, it shows that the main jet outer layer can trace the deviation due to the { collision} between the flows. The fourth and fifth panels present the higher velocities, between 10 and 13 km~s$^{-1}$, and 13 to 20 km~s$^{-1}$, respectively. The emission is stronger in the leading part, meaning a significant expansion exists.}
    \label{fig:velchanel}
\end{figure*}

To predict the observation of this kind of jet collision, we have made emission maps and integrated them over the 'z' axis, which we have considered as the line of sight. That is, we are assuming that the main flow of the primary jet propagates along the 'y' axis and the main flow of the secondary jet propagates along the 'x' axis. Therefore, by integrating over the 'z' axis, we can analyze the formation of the hole in the envelope of our object. To get the molecular emission, we consider that all the gas has a fraction of CO concerning the total density of $\chi_{CO}=1.67 \times 10^{-4}$. To find the emission, we use equation 3 of \citet{Rodriguez-Gonzalez2023}, 

\begin{equation}
 {j_{\rm CO ({2 \to 1})}}=\frac{1}{4\pi}\frac{g_1}{Z(T)}n_{\rm CO} \cdot e^{\frac{-E_{\rm {2\to1}}}{k T_{{2\to1}}}} A_{\rm {2\to 1}} E_{\rm {2\to 1}},\\
\end{equation}
where, $g_1=3$ is a degeneracy factor, $Z(T)=\sum^N_i e^{-T_{\rm lev CO}/T(i,j)}$ is the partition function, with $T_{\rm lev CO}$ as the temperature of the corresponding energy levels,
  $A_{{2\to1}}=7.16 \times 10^{-7}$ s$^{-1}$.
is the spontaneous emission coefficient,  $E_{{2\to1}}=h \nu_{{2\to1}}$ is the energy of the transition
with,  $\nu_{{2\to1}}=230.538$ GHz and $h$ is the
Planck constant \citep{Rybicki}. We have assumed that $n_{CO} = \chi_{CO}*n_g$, where, $n_g$ is the numerical gas density. 

Figure~\ref{fig:velchanel} normalized CO Emission map for velocities 1 < v < 20 km~s$^{-1}$, in the direction of the observer, considering the jet is travelling along the plane of the sky. In the upper panel (between 1 and 4 km /s), we show the emission delineates the outer cavity walls, including the bubble generated by the secondary jet in the bottom left corner. A second bubble can be noticed by the lack of emission at a distance of 2 $\times$ 10$^4$ au, corresponding to the broadest part of the outflow. Also, the leading bow shock shows no emission, which can be explained by the fact that it is expanding at larger velocities. The second panel (between 4 and 7 km~s$^{-1}$) shows the inner cavity walls, while the emission is spread more evenly across the outflow. Again, the leading shock, like the secondary bubble, shows no emission in this velocity range. The third panel (between 7 and 10 km~s$^{-1}$) shows more emission inside both bubbles, and under $x = 10^4$ au, the main jet outer layer can trace the deviation due to the { collision} between the { momentum of the} flows. In the fourth and fifth panels (between 10 and 13 km~s$^{-1}$ and between 13 and 20 km~s$^{-1}$), the emission is stronger in the leading part, which means there is a significant expansion, probably due to the bullets catching up with each other in a previous simulation time. That is comparable to Cep E, where the outflow southern lobe appears filled with material connecting the jet and the cavity. { This could explain the hole observed in the southern cavity of Cep E in the low velocity CO ($2\to1$) emission. Even when other mechanisms to change the CO emission remain to be explored, this scenario is a plausible explanation for the observed maps.}

\section{Conclusions}
\label{Sec:5}
We have analyzed the {collision} of a system of two outflows produced by the collimated jets ejected from two protostars, taking some of the physical constraints imposed by the protostellar binary system Cep\, E and we have carried out a 3D hydrodynamic numerical model. The free parameters prevent a fine-tuned model, but we have reproduced the physical signatures associated with this complex { collision} in { Cepheus E}, where the momentum in all three spatial directions and the ram pressure can significantly influence the structure morphology and dynamical evolution of the primary envelope, { driven by the protostar Cep-A}. { To quantify the interaction between Cep E outflows}, we have produced a fiducial model consisting of a single source ejecting a collimated hypersonic jet that propagates in a density stratified environment in the $y$ axis, which we have named the primary jet. Then, we produced a second model that includes the fiducial parameters. After 660 yr, a secondary source ejects a collimated jet that moves parallel to the $x$ axis from $x=1500$ au $y=1500$ and $z=250$ au, to avoid direct collision between both simulated jets. {Nevertheless, since they expand perpendicularly to their main axis, they affect each other's dynamics and the morphology of their cavities}. We summarize our conclusions in the following points: 

\begin{itemize}
    \item {We have estimated the probability of an encounter between different outflow components, which are $\sim80\%$ for a cavity-cavity encounter and $\sim 20\%$ for a jet-jet encounter.}
    \item The fiducial model shows that the primary high-velocity jet produces a low-velocity cavity comparable to axisymmetric simulations such as Raga et al. (2014), and \cite{Rivera-Ortiz2023}.
    \item The propagation of a secondary high-velocity jet inside the cavity generated by the primary jet induces perturbations inside the primary cavity, showing a morphology of density structure associated with the first interaction. This { collision} creates an intermediate velocity bubble that travels behind the primary jet's leading bowshock it tends to expand and feedback the primary outflow cavity. 
    \item The primary jet morphology also gets affected. Compared with the fiducial model and axisymmetric simulations, the primary jet gets wider faster by interacting with the secondary jet. Then, the internal knots produced by the variability in the primary jet reach a larger size that is approximately twice the unperturbed knots, and therefore, its density is reduced considerably. This can also explain why the M2 model at $1875$ yr is shorter than M1 at the same age, since the dynamical friction depends on the surface of the propagating knots.
    \item The secondary jet produces a shocked region inside the primary cavity wall, deeply affecting the initial morphology. This can be observed as a stationary bubble aligned to the secondary jet propagating axis where the density and temperature increase significantly. This { could} explain the hole observed in the southern cavity of Cep E in the low velocity CO ($2\to1$) emission.
    \item The { collision} region of the jets, the zone where both jets are at a minimum distance from each other, is not significantly affected compared to the dynamics and morphology of the main cavity. This should depend on the impact parameter between both jets, which is a parameter that remains to be explored in more detail in future simulations.
    \item { For our simulation, the primary jet exhibits a pseudo-prescesion produced by the outflows interaction}. In the projected plane $xy$ where both jets are propagating, the primary jet gets a maximum deviation of $\sim 3.5^\circ$ from its initial direction. This could explain why the northern jet in Cep E exhibits precession around its main propagating axis, the southern jet's main axis is not aligned with the northern lobe, and its precession amplitude is also different from that of its northern counterpart.
    \item Given its density, radius and velocity, the main jet injects mass with a rate of $4.3\times 10^{-8}$M$_\odot$ yr$^{-1}$. Then, we have calculated the mass of each of the moving knots in models M1 and M2 at $t=1875$ yr, which are, in both cases, around $5\times 10^{-5}$M$_\odot$, that would correspond to an injection mass rate of $\sim 1.3 \times 10^{-7}$ and is around 3 times the previous value for jet injection rate. This factor agrees with the entrainment effect discussed by \cite{Rivera-Ortiz2023}.
\end{itemize}

 Since the interaction between the two supersonic outflows occurs at { thousands of au} from the protostars, we did not consider some elements that are likely to be important for the evolution of the protostellar flows:
\begin{itemize}
    \item the disks surrounding the protostellar objects, 
    \item radiative transfer,
    \item magnetic fields,
    \item gravitational forces of the binary system,
\end{itemize}

 In the particular case of Cepheus E, the collimated flows (or jets) from the north and south exhibit similarities in phase and period of the jet injection but not in spatial frequency.   Then, our analysis focuses on the purely hydrodynamic effects of the interaction between two outflows.
Therefore, there is a large range of possibilities
for future studies of the molecular outflows ejected by protostellar Class 0 objects.



\section*{Acknowledgements}

The authors are grateful to B. Lefloch for his participation and valuable contributions to the development of this article, and appreciate the comments of the anonymous referee, who helped improve the content of this work. We acknowledge the support of the UNAM-PAPIIT grants IN110722, IN102724, and AG101125, as well as the Miztli-UNAM supercomputer project LANCAD-UNAM-DGTIC-123 (2022-1) and LANCAD-UNAM-DGTIC-128 (2023-1).
E.A. and A.D. are grateful for the grants, CONAHCYT numbers 4050143 and 
825308, respectively.

\end{document}